# A Method for Neutron-Gamma Pulse Shape Discrimination of CLYC Detector Based on a Gated Residual-Linear Attention Network

Shengduo Liu [a], Shiwei Jing [a, *], Weiyang Zhang [a], Jia Song [a], Sijia Zhou [a], Hailong Xu [a], Yue Sun [a], Zebin Li [a], Yuxuan Gu [a], Siqi Liu [a], Tian Zhang [a], Zhihua Gao [a], Guofeng Qu [a, *], Fuquan Jia [b]

[a] School of Physics, Northeast Normal University, Changchun 130024, Jilin, China

[b] Jilin Jianzhu University, Changchun 130118, China

*Corresponding author.

E-mail address: jingsw504@nenu.edu.cn (Shiwei Jing), quguofeng@scu.edu.cn (Guofeng Qu)

**Abstract:** The discrimination of neutron and gamma pulse shapes is a key technology in fields such as nuclear safety monitoring and radiation assessment. An enhanced recursive gated cyclic residual-sparse linear attention network is developed on the CLYC detector experimental platform to overcome weak noise resistance, limited feature extraction and inferior real-time performance of conventional algorithms. The experimental dataset comprises 19,971 samples, which were pre-processed and stratified for model training and testing. Results indicate that the proposed algorithm achieves a quality factor of 2.2, with a classification accuracy of 98.7% and a recall rate of 99.4%. It achieves an accuracy of 95.1% under the 20 dB low signal-to-noise ratio condition, exhibiting excellent anti-noise ability.With around 2.8 million parameters, the model takes merely 0.05 ms to process a single pulse on GPU, satisfying real-time monitoring and embedded deployment demands.

## 1 Introduction

With the rapid development of nuclear energy and technology, nuclear safety has become an increasingly important concern. Effective and accurate detection of nuclear radiation is a prerequisite for nuclear safety. Neutrons have high penetrating power and are highly hazardous to the human body, which imposes stricter requirements on the use of neutron sources and the measurement of neutron radioactivity [1]. When detecting neutrons, neutron radiation is typically accompanied by gamma rays, and the presence of gamma rays can introduce interference into subsequent data analysis and processing. Consequently, research into methods for distinguishing between neutrons and gamma rays is of great significance for neutron measurement [2].

The Pulse Shape Discrimination (PSD) method is the most widely applied technical approach in the field of n/γ discrimination. Over the past several decades, n/γ discrimination technology has evolved from early analogue circuit-based discrimination and processing to the implementation of discrimination algorithms using digital systems [3]. With the advancement of digital technology, traditional analogue time-domain discrimination methods have gradually shifted towards digitalization. In 2007, Mellow et al. first proposed the pulse gradient method based on digital PSD [4]. In recent years, digital discrimination methods have undergone further development, with intelligent algorithms based on pattern recognition, machine learning and deep learning driving the rapid advancement of intelligent pulse shape discrimination methods [5]. In 2020, Song Haisheng et al. applied a back-propagation (BP) neural network to particle discrimination in mixed n/γ-ray fields. The results demonstrated that the BP neural network discrimination method not only achieves effective discrimination but also improves discrimination time [6]. In the same year, Griffiths et al. proposed a method for n/γ pulse shape discrimination based on convolutional neural networks [7]. In 2021, Ma Tongda achieved efficient n/γ discrimination by integrating principal component analysis, genetic algorithms and support vector machines [8]. In 2022, Lu et al. utilised a combination of CLYC detectors and SiPMs to propose a convolutional neural network-based n/γ discrimination scheme using a two-dimensional matrix as input [9]. In 2025, Guo et al. proposed a method for neutron-gamma discrimination in fast signals from SiPM arrays based on deep neural network optimization, achieving a discrimination probability of 98.8% [10].

Pulse shape discrimination methods based on intelligent algorithms have become the prevailing trend in the field of n/γ discrimination. Although traditional time-domain PSD methods can achieve good discrimination results under specific conditions, they are relatively sensitive to noise and are difficult to apply to discrimination tasks in low-activity neutron-gamma mixed radiation fields [11]. Although frequency-domain PSD methods possess strong noise immunity, they impose stringent requirements on analogue-to-digital conversion and sampling hardware, typically requiring a sampling rate of over 1 GS/s to ensure ideal discrimination performance [12]. In contrast, intelligent PSD methods are not only insensitive to noise but also significantly reduce the hardware requirements for acquisition cards, whilst being capable of extracting multi-dimensional pulse features that are difficult to obtain using traditional time-domain and frequency-domain methods [13].

# 2 Materials and methods

## 2.1 n/γ Pulse Shape Discrimination Algorithm Based on RGLR-SLA

An optimized Recurrent Gated Loop Residual-Sparse Linear Attention (RGLR-SLA) network is presented to overcome limitations of conventional algorithms based on hierarchical deep feature learning. This network captures multi-level shape information of pulses through multi-scale feature perception, utilizes the synergistic fusion of local and global features to extract key discriminative features, and incorporates an efficient attention mechanism to enhance computational efficiency, thereby achieving high-precision, high-robustness, and high-real-time n/γ pulse classification.

### 2.1.1 Core Principles of the Algorithm

The RGLR-SLA network achieves high-precision classification of neutron-gamma pulses through multi-scale feature detection, the synergistic fusion of local and global features, and efficient sparse linear attention mechanisms.

Multi-scale feature detection utilizes parallel convolutional channels to separately extract nanosecond-level rising edge details, microsecond-level decay rates, and 10-microsecond-level overall contour features, comprehensively characterizing the physical differences between neutron and gamma pulses across different time scales, thereby addressing the limitation of traditional algorithms that can only capture information at a single scale [14]. Local-global feature collaborative fusion employs deep separable convolutions to extract local waveform details, combines sparse linear attention to model long-range temporal dependencies, and utilizes a learnable gating mechanism for adaptive weighted fusion. This approach enhances global features to suppress interference in high-noise environments and highlights local features to improve resolution in low-noise conditions [15]. Efficient sparse linear attention reduces the computational complexity of traditional self-attention from $O(n^2)$ to $O(n)$. By combining sparsification strategies with linear projection, it significantly improves computational efficiency while preserving the ability to model global features, thereby meeting the real-time processing requirements for long pulse sequences [16].

### 2.1.2 Algorithm Implementation and Model Structure

The RGLR-SLA network is implemented using the PyTorch framework. The overall model architecture consists of an input layer, a multi-scale block embedding module, an enhanced RGLR module stack, an enhanced SLA module, an adaptive pooling layer, and a classifier. The complete architecture is shown in **Figure 1**.

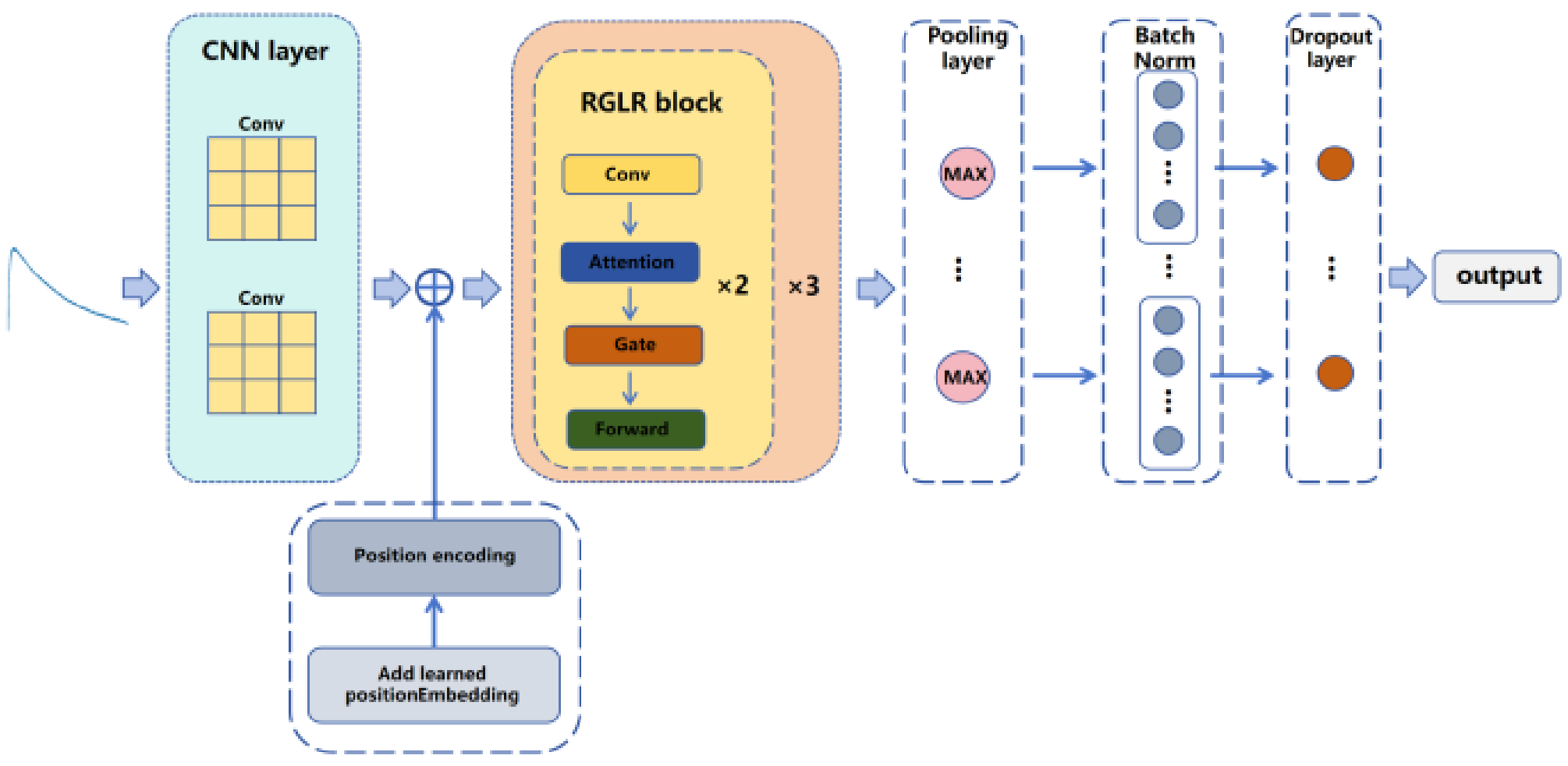

**Fig. 1** RGLR-SLA Network architecture diagram

The input to the input layer is a preprocessed single-channel one-dimensional impulse sequence with a length of 65536 samples, where each sample represents a normalized voltage value.The Multi Scale Patch Embed module converts the raw pulse sequence into a multi-scale feature map, providing a foundation for subsequent feature learning. The module contains three parallel convolutional paths, each designed to extract features at different temporal scales, as shown in **Table 1**.

The Enhanced RGLR module consists of four stacked layers, including layer normalization, a local feature path, a global feature path, gated fusion, and residual connections. The local path uses two layers of depth-separable convolutions to extract local waveform features; the global path employs bidirectional GRUs to capture long-range temporal dependencies; a learnable gating mechanism adaptively weights and fuses local and global features, while residual connections mitigate the vanishing gradient problem and enhance training stability.

The Enhanced Sparse Linear Attention (Enhanced SLA) module obtains query, key, and value vectors through linear projection. It employs a Top-k sparsity strategy to reduce computational complexity and combines linear attention to achieve efficient global feature modeling. While retaining the ability to model long-range dependencies, it reduces computational complexity from $O(n^2)$ to $O(n)$.

**Table 1** Multiscale embedded module structural parameters

| | Convolutional layer | Convolution kernel Size | Step | Number of input channels | Number of output channels | Activation function |
|---|---|---|---|---|---|---|
| Channel 1 (small-scale features) | 1D conv layer | 64 | 64 | 1 | 32 | GELU |
| Channel 2 (mesoscale features) | 1D conv layer | 128 | 64 | 1 | 32 | GELU |
| Channel 3 (Large-scale features) | 1D conv layer | 256 | 64 | 1 | 32 | GELU |

The Adaptive Pooling layer combines attention-weighted average pooling with max pooling to convert variable-length feature sequences into fixed-dimensional 128-dimensional feature vectors. The classifier consists of two fully connected layers and uses temperature-scaled Softmax to output neutron and gamma classification probabilities, enabling end-to-end pulse discrimination.

The Classifier module is responsible for mapping the fixed-dimensional feature vectors to classification results. The specific structure of the Classifier module is shown in **Table 2**. The output dimensions are (Batch_size, 2), corresponding to the gamma and neutron probabilities, respectively.

**Table 2** Fully connected layer structure of a classifier

| | Input dimensions | Output dimensions | Activation function | dropout |
|---|---|---|---|---|
| Fully Connected Layer 1 | 128 | 64 | 64 | 0.1 |
| Fully Connected Layer 2 | 64 | 2 | - | - |

The foregoing content elaborates on the parameters and functional mechanisms of each module in the RGLR-SLA network.The complete end-to-end workflow of the proposed model is systematically presented in **Figure 2**, which clearly illustrates the integration of multi-scale embedding, gated residual fusion, and sparse linear attention throughout the pulse discrimination process.

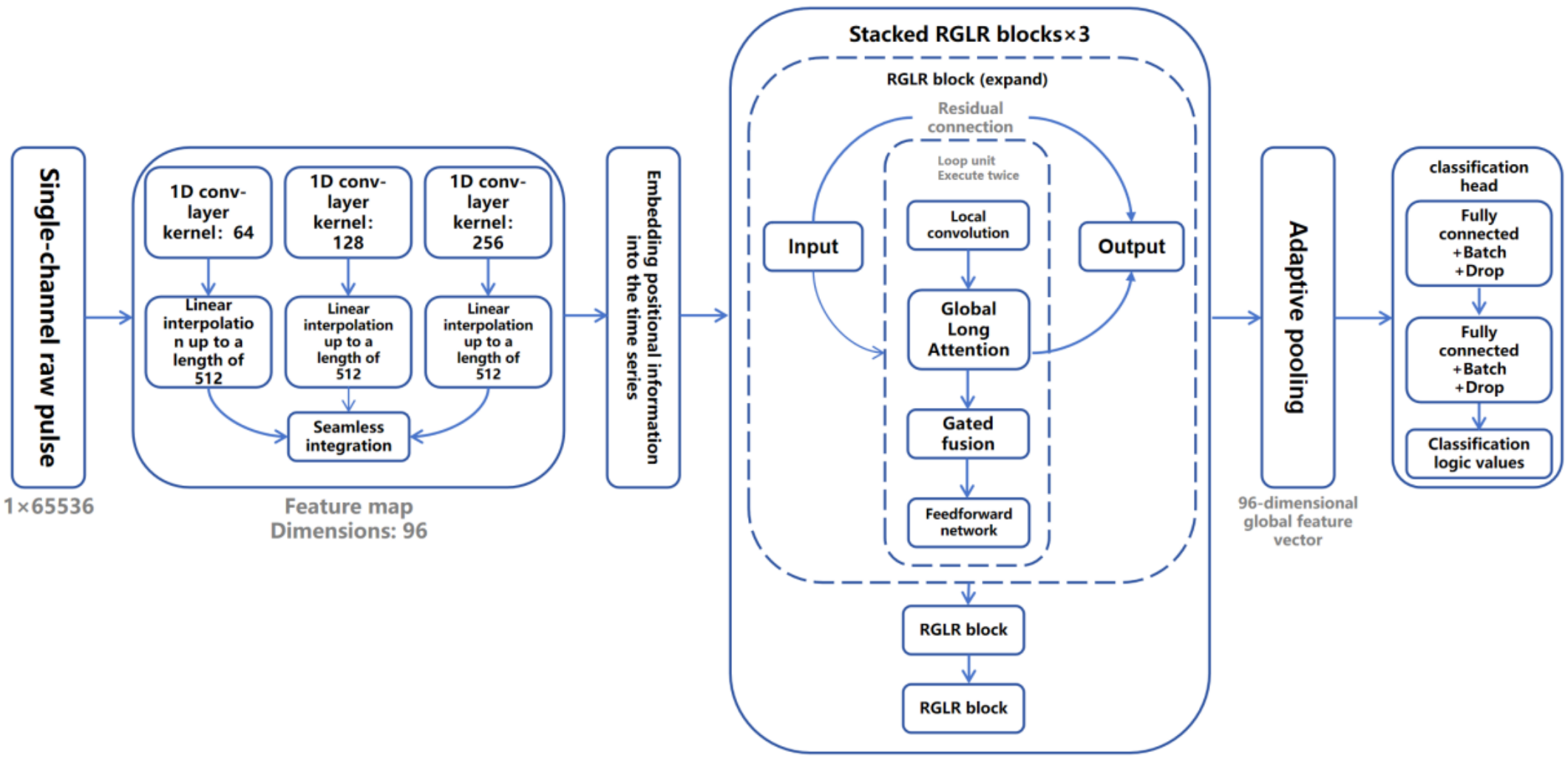

**Fig. 2** Process Flowchart for the RGLR-SLA Model

## 2.2 Experimental Platform

The quality of the pulse data directly determines the effectiveness of algorithm training and its generalization capability. To obtain authentic neutron and gamma pulse signals, an experimental platform was established comprising a radiation source, a CLYC detector, and a signal amplification and high-speed data acquisition system. The radiation source selected was the NG-9 D-D neutron generator developed by Northeast Normal University, which can produce 2.5 MeV neutrons and create a mixed radiation field environment; In addition, $^{137}$Cs and $^{60}$Co standard sources were selected to acquire pure gamma pulses for verifying the accuracy of the discrimination algorithm [17].

The CLYC detector consists of a 25 mm × 25 mm crystal with a $Ce^{3+}$ doping concentration of 0.5 mol% and a $^{6}$Li enrichment of ≥95%, paired with a photomultiplier tube (PMT) as the optical-to-electrical signal converter. The weak electrical signal output by the PMT is amplified by a preamplifier and transmitted via coaxial cable to a high-speed data acquisition board. The acquisition board employs a rising-edge trigger to prevent false triggering caused by noise. The experimental setup is shown in **Figure 3**.

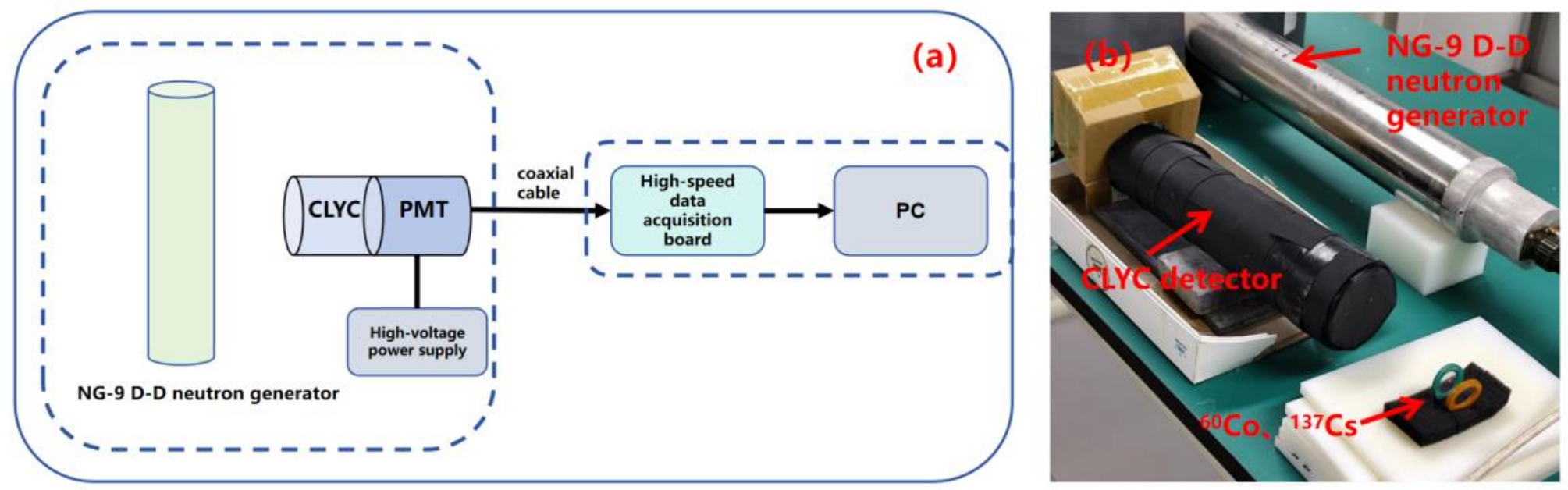

**Fig. 3** Schematic diagram and photograph of the experimental setup: (a) Schematic diagram; (b) Actual product image

## 2.3 Data Preprocessing

To eliminate the interference caused by data fluctuations on the classification algorithm and ensure the consistency of the input signals, the data must be pre-processed. The pre-processing workflow follows a logical sequence comprising noise suppression, amplitude normalisation and dataset partitioning, as illustrated in **Figure 4** [18]..

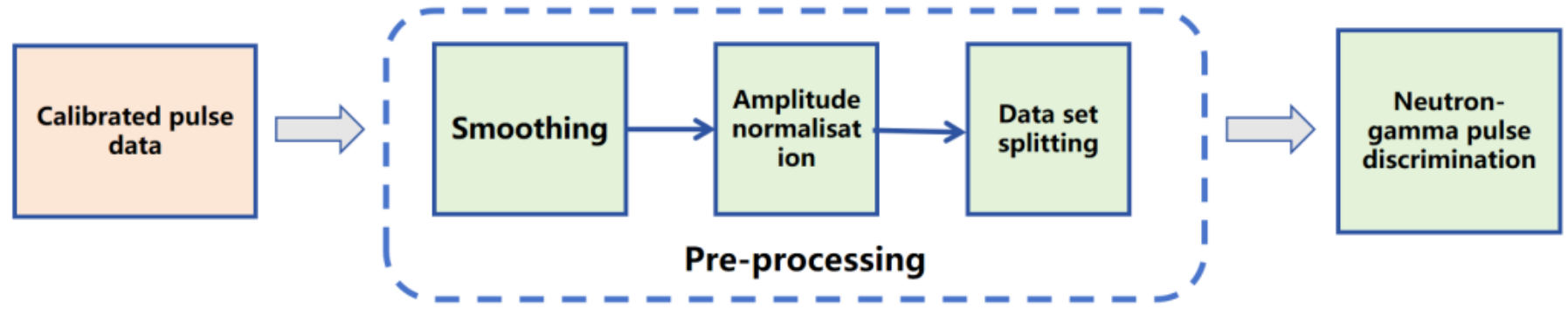

**Fig. 4** Pre-processing workflow for the identification of neutron and gamma-ray pulses

After preprocessing, the data must be validated against three criteria—pulse integrity, signal-to-noise ratio (SNR), and shape consistency—to ensure it is suitable for training. All pulses must retain their complete rising edges, peaks, and decay phases, with no truncation or distortion; SNR is calculated using the formula SNR = peak amplitude / baseline standard deviation. The SNR of all samples must be greater than or equal to 30 dB to meet the basic signal quality requirements of the discrimination algorithm. Regarding shape consistency, pulses from the same particle type exhibit stable shape characteristics after normalization; for example, neutron pulses have a gentle rise and slow decay, while gamma pulses have a steep rise and rapid decay. Only data that has passed quality verification can serve as reliable data for subsequent feature extraction and classification [19].

A preprocessed clean pulse dataset with 19,971 valid samples is employed for model training and testing. Among these, there are 8,974 neutron pulses (44.9%) and 10,977 gamma pulses (55.1%), with a generally balanced distribution of sample categories and no significant bias. To avoid model overfitting and ensure the objectivity of the evaluation results, a stratified random sampling method was used to divide the dataset into training, validation, and test sets in a 75:15:10 ratio. The specific division results are shown in **Table 3**.

**Table 3** Results of the dataset split

| | Neutron | Gamma | Total |
|---|---|---|---|
| training set | 6731 | 8269 | 15000 |
| Validation set | 1346 | 1654 | 3000 |
| test set | 897 | 1074 | 1971 |

Stratified sampling ensures that the ratio of neutron to gamma events within the three datasets matches that of the original dataset, thereby avoiding evaluation biases caused by differences in sample distribution [20].

# 3 Experiments and Discussion

## 3.1 Evaluation criteria

In this experiment, the results were evaluated using the following metrics: Figure of Merit (FoM), Accuracy, Precision, Recall, and F1 score [21].

(1) Figure of Merit

Used to comprehensively measure classification performance, it is defined as:

$$FoM = (\mu_n - \mu_\gamma) / (FWHM_n + FWHM_\gamma) \quad (1)$$

where $\mu_n$ and $\mu_\gamma$ are the peak positions of the Gaussian functions fitted to the neutron and gamma pulse discrimination factors, respectively, and FWHMn and FWHMγ are the corresponding full widths at half maximum.

(2) Accuracy

Used to measure the model’s overall discrimination capability, defined as:

$$\text{Accuracy} = (\text{number of true positives} + \text{number of true negatives}) / \text{Total sample size} \times 100\% \quad (2)$$

where the number of true positives is the number of correctly identified neutron pulses, and the number of true negatives is the number of correctly identified gamma pulses.

(3) Precision

Reflects the proportion of true neutron pulses among those identified by the model, defined as:

$$\text{Precision} = \text{number of true positives} / (\text{number of true positives} + \text{number of false positives}) \times 100\% \quad (3)$$

(4) Recall

Reflects the model's ability to capture true neutron pulses, defined as:

$$\text{Recall} = \text{number of true positives} / (\text{number of true positives} + \text{number of false negatives}) \times 100\% \quad (4)$$

(5) F1 score

A comprehensive evaluation metric combining precision and recall, defined as:

$$\text{F1} = 2 \times (\text{Precision} \times \text{Recall}) / (\text{Precision} + \text{Recall}) \quad (5)$$

## 3.2 Model Ablation Experiments

To validate the effectiveness of each core module in the RGLR-SLA network and quantify the contributions of multi-scale embedding, recursive gated loop residuals, and sparse linear attention to classification performance, a series of ablation experiments are conducted to verify the effectiveness of each core module.

### 3.2.1 Validation of the Multiscale Block Embedding Module

Multi-scale embedding serves as the first-level feature extraction module in RGLR-SLA, capturing micro-, meso-, and macro-scale features of the pulse through three parallel convolutional channels. To validate the necessity of the three-scale design and compare the performance differences among single-scale, two-scale, and three-scale approaches, we designed seven convolutional channels with different scales and configurations for comparative experiments. The experimental results are shown in **Table 4**.

The distinguishing features of different energy pulses are distributed across different time scales; for example, the rising edge is more critical for low-energy pulses, while the slow decay component is more critical for high-energy pulses. Therefore, single-scale models perform well within specific energy ranges but underperform in others. The three-scale configuration achieves feature fusion through $1\times1$ convolutions after channel concatenation, enabling the network to adaptively select the critical scale. In the low-energy region, the network automatically boosts the weights of the 64-kernel channel to focus on the details of the rising edge; in the high-energy region, it boosts the weights of the 256-kernel channel to focus on the slow decay trend. The FoM improves from 1.42–1.55 in the single-scale model to 2.20 in the three-scale model, representing an increase of 42.4%–55.6%.

**Table 4** Results of multi-scale embedding and ablation experiments

| Specifications | Convolution kernel | FoM | Accuracy(%) | Accuracy in the low-energy region(%) | Accuracy in the high-energy region(%) |
|---|---|---|---|---|---|
| Single-scale-S | 64cores | 1.55 | 92.3 | 78.5 | 94.2 |
| Single-scale-M | 128cores | 1.48 | 91.8 | 85.6 | 89.3 |
| Single-scale-L | 256cores | 1.42 | 90.5 | 88.2 | 85.7 |
| Dual-scale-SM | 64+128cores | 1.78 | 95.6 | 89.4 | 96.8 |
| Dual-scale-SL | 64+256cores | 1.72 | 94.8 | 86.3 | 95.5 |
| Dual-scale-ML | 128+256cores | 1.68 | 94.2 | 90.1 | 93.6 |
| Tri-scale-SML | 64+128+256 cores | 2.2 | 98.7 | 96.5 | 99.1 |

### 3.2.2 Verification of the Local-Global Fusion Mechanism in the RGLR Module

The RGLR module is the core of RGLR-SLA. It achieves feature fusion through parallel processing of local convolutional paths and global GRU paths, combined with adaptive gating. To verify the contributions of each component, the effectiveness of the RGLR module is demonstrated in terms of FoM and accuracy. The experimental results are shown in **Table 5**.

**Table 5** Results of the RGLR module ablation experiment

| Specifications | Local path | FoM | Accuracy(%) | SNR=30dB Accuracy(%) | SNR=20dB Accuracy(%) |
|---|---|---|---|---|---|
| CNN | 3-layer standard convolutional | 1.65 | 93.5 | 89.2 | 76.8 |
| Local-Only | 2-layeDeep separable convolutions | 1.72 | 94.8 | 91.5 | 81.3 |
| Global-Only | - | 1.8 | 95.6 | 93.8 | 85.6 |
| RGLR-Concat | Deep separable convolutions | 1.95 | 96.8 | 95.2 | 89.4 |
| RGLR-Add | Deep separable convolutions | 2.02 | 97.5 | 96.1 | 91.8 |
| RGLR-Fixed | Deep separable convolutions | 2.08 | 97.9 | 96.8 | 93.5 |
| RGLR-Adaptive | Deep separable convolutions | 2.2 | 98.7 | 98.2 | 95.8 |

The pure local path performs well under clean signals but drops sharply to 81.3% under strong noise. In contrast, the pure global path is more robust under strong noise but underperforms the local path under clean signals. The fusion of the two achieves performance optimization across the entire noise spectrum. Simple concatenation leads to feature redundancy; while the number of parameters increases, the FoM remains at only 1.95. Element-wise addition enables feature interaction, boosting the FoM to 2.02. Adaptive gating dynamically adjusts local/global weights based on the input, further improving FoM to 2.2, with the most significant advantage observed under strong noise. The local path of RGLR employs deep separable convolutions, which reduce the number of parameters by approximately 8 times and computational complexity by approximately 4 times compared to standard convolutions, while maintaining comparable feature extraction capabilities. This is key to RGLR-SLA maintaining a low number of parameters.

#### 3.2.3 Verification of the efficiency-accuracy trade-off in Sparse Linear Attention (SLA)

The SLA module is used to further refine global features and establish long-range dependencies. The computational complexity of self-attention in a standard Transformer is $O(n^2)$; for a sequence length of 1024, the computational load is $O(1024^2)=O(1M)$, which makes it difficult to meet real-time requirements. SLA reduces the complexity to $O(n)$ through sparsification and linear projection. This subsection designs attention mechanisms with different configurations for comparison to demonstrate the efficiency and accuracy of the SLA module. Quantitative results from ablation experiments on different attention mechanisms are summarized in **Table 6**.

**Table 6** Results of attention mechanism ablation experiments

| Specifications | Types of attention | FoM | Accuracy(%) | GPU time (ms) | CPU time (ms)） |
|---|---|---|---|---|---|
| No-Attention | No attention, pooling | 2.05 | 97.2 | 0.03 | 0.25 |
| Full-Attention | Standard self-attention | 2.28 | 99.1 | 2.5 | 15.8 |
| Liner-Only | Linear attention (without sparsification) | 2.15 | 98.2 | 0.12 | 0.85 |
| Sparse-Only | Sparse Attention (Non-linear projection) | 2.18 | 98.5 | 0.08 | 0.62 |
| SLA | Sparse + Linear Attention | 2.2 | 98.7 | 0.05 | 0.3 |

The FoM of the No-Attention configuration is only 2.05, significantly lower than that of all attention configurations. This demonstrates that modeling long-range dependencies is crucial for classification performance, particularly for capturing global trends with slow decay components. Standard self-attention achieves the highest FoM, but its processing time reaches 2.5 ms, with a theoretical throughput of only 63 cps, far below the real-time requirements of nuclear radiation monitoring. SLA suffers only a 3.5% loss in FoM compared to Full-Attention, yet its processing speed is 50 times faster. This excellent trade-off between efficiency and accuracy enables RGLR-SLA to meet real-time processing demands. Comparing Linear-Only and Sparse-Only, their FoM values are similar, but Sparse-Only is faster. SLA combines the strengths of both approaches: it reduces computational load through sparsification and lowers dimensionality via linear projection, achieving dual optimization of speed and accuracy.

The ablation results indicate that the performance of the RGLR-SLA network depends on the combined effects of its core modules. The multi-scale block embedding comprehensively captures the cross-scale physical characteristics of the pulse, achieving a maximum improvement in the FoM of 55.6%; the enhanced RGLR module significantly improves classification accuracy under low signal-to-noise ratio conditions through adaptive fusion of local and global features; and Sparse Linear Attention (SLA) improves computational efficiency by 50 times with less than 3% loss in accuracy, meeting real-time processing requirements.

## 3.3 Model Training and Results

The model uses a binary cross-entropy loss function. The dataset has a balanced class distribution, with no class weights applied, and label smoothing is introduced to improve generalization. The AdamW optimizer is used with a weight decay of $10^{-4}$. The learning rate is managed using the OneCycleLR strategy, which consists of three phases—warm-up, stabilization, and cooling—with a total of 30 training epochs. The training process was accelerated using progressive dropout, gradient clipping, and Auto-Mixed Precision (AMP), combined with an early stopping mechanism to prevent overfitting.

With a batch size of 32 and an RTX 4070 laptop GPU, the model converged rapidly after approximately 2.5 hours of training. The training loss decreased from 0.68 to 0.03, and the validation loss decreased from 0.65 to 0.04. The difference between training and validation losses was less than 0.01, indicating no significant overfitting. The final training accuracy reached 99.1%, and the validation accuracy was 98.9%, demonstrating stable convergence and good generalization performance. The trends of training and validation losses over epochs are shown in Figure 5.

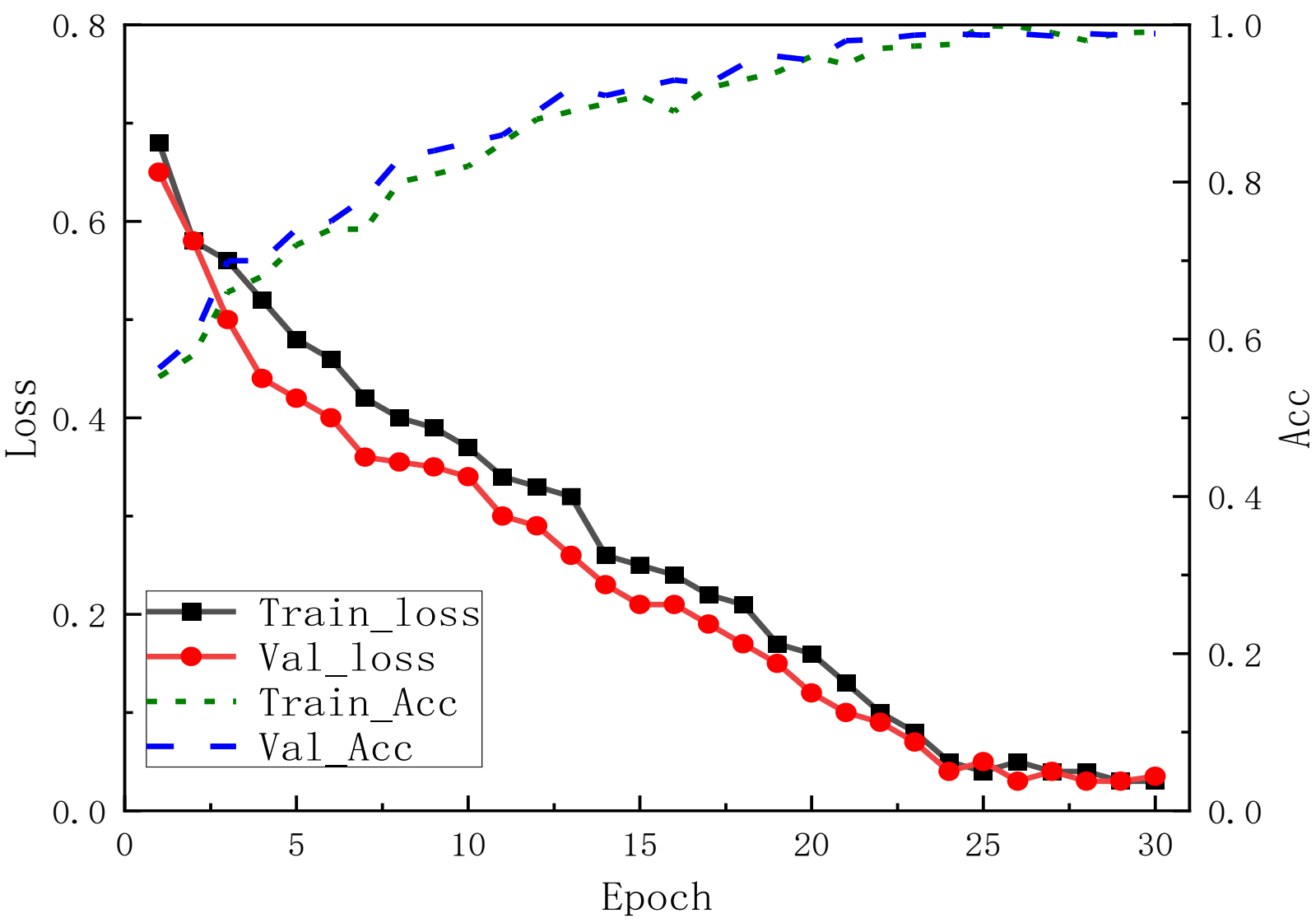

**Fig. 5** Loss values and Accuracy during the RGLR-SLA network training process

## 3.4 Comparison of Classification Performance

Classification performance is quantified using five metrics: quality factor, accuracy, precision, recall, and F1 score. The comparison results for each metric are shown in **Table 7**. The RGLR-SLA network proposed outperforms all others across all five metrics. Its quality factor reaches 2.2, representing a 37.5% improvement over the best-performing traditional method—the frequency gradient method; the classification accuracy reaches 98.7%, demonstrating the model's strong ability to distinguish between neutron and gamma pulses.

**Table 7** Comparison of performance metrics for different pulse discrimination algorithms

| | FoM | Acc | Precesion | Recall | F1 Score |
|---|---|---|---|---|---|
| Charge comparison method | 1.2 | 91.4% | 85.4% | 86.7% | 68.2% |
| Rising-time method | 1.5 | 93.5% | 88.2% | 83.3% | 85.7% |
| Frequency gradient method | 1.6 | 94% | 89.4% | 91.1% | 90.2% |
| CNN+LSTM+Transformer | 1.8 | 96.4% | 93.5% | 96% | 94.7% |
| RGLR-SLA | 2.2 | 98.7% | 96.4% | 99.4% | 97.9% |

## 3.5 Analysis of noise immunity

In practice, the noise in neutron-gamma mixed radiation fields is complex. To test the algorithm's robustness, Gaussian white noise of varying intensities was added to the test set pulses to simulate changes in signal-to-noise ratio ranging from 40 dB to 15 dB under high-noise conditions. The experimental results are shown in Figure 6. As the SNR decreases, the performance of all methods deteriorates; however, the RGLR-SLA network exhibits the slowest decline. When the SNR drops to 20 dB, the accuracy of traditional methods falls below 80%, whereas the algorithm proposed in this paper maintains an accuracy of 95.1%. This indicates that the RGLR-SLA network has learned intrinsic shape features for classification through training and possesses excellent noise resistance.

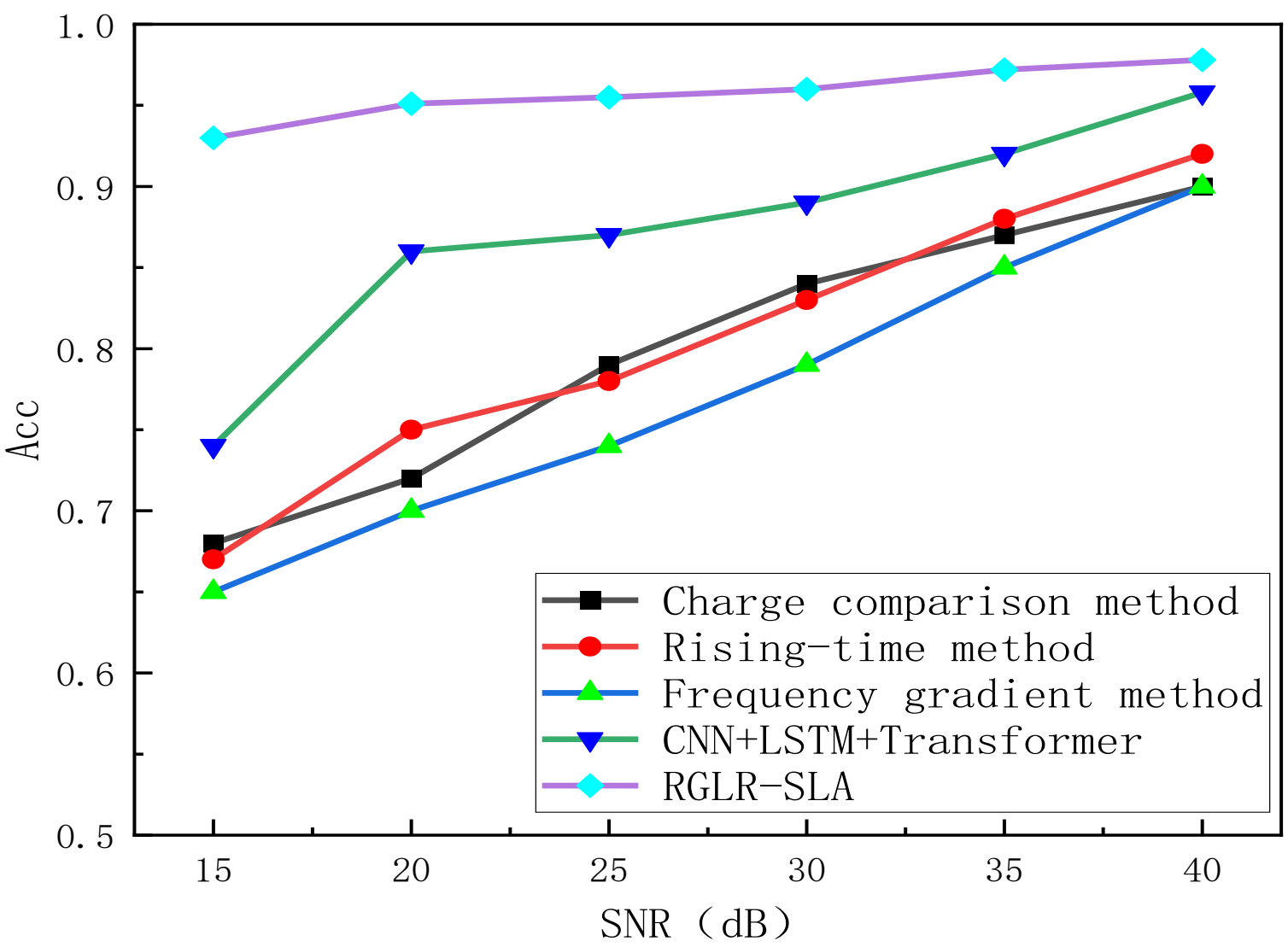

**Fig. 6** Variations in the accuracy of different classification algorithms at different signal-to-noise ratios

### 3.6 Real-Time Analysis

The processing speed of the algorithm is critical for engineering applications. The RGLR-SLA model has approximately 2.8 million parameters. On an Intel Core i7-12800HX CPU platform, the average processing time per pulse is approximately 0.3 ms; on an NVIDIA RTX 4070 Laptop GPU platform, this can be further reduced to 0.05 ms. Even in embedded systems using only a CPU, the theoretical processing throughput meets the real-time requirements of radiation monitoring systems.

## 4 Conclusion

This work targets the drawbacks of poor noise robustness, limited feature extraction, and insufficient real-time performance in traditional methods for neutron-gamma pulse shape discrimination. A platform for acquiring experimental data has been established based on a CLYC detector, and an enhanced Recursive Gated Recurrent Residual-Sparse Linear Attention Network is proposed. This network employs a hierarchical architecture combining multi-scale block embedding, recursive gated recurrent residual modules and sparse linear attention, enabling the collaborative learning of micro-level details, meso-level structures and macro-level contours of pulse signals, as well as the adaptive fusion of local waveform and global temporal dependencies. Ablation experiments validate the critical role of multi-scale feature extraction, local-global adaptive gated fusion, and sparse linear attention in performance enhancement, with the modules working in concert to achieve a balance between accuracy and efficiency.

Experimental results demonstrate that the proposed RGLR-SLA network achieves a FoM of 2.2 on real-world pulse datasets, with a classification accuracy of 98.7%, precision of 96.4%, recall of 99.4%, and an F1 score of 97.9%. Its overall classification performance significantly outperforms traditional algorithms such as the charge comparison method, rise time method, and frequency gradient method, as well as benchmark deep learning models. Even in highly noisy environments with a signal-to-noise ratio as low as 20 dB, the algorithm maintains a classification accuracy of 95.1%, demonstrating excellent robustness against interference. Furthermore, with approximately 2.8 million model parameters, the single-pulse processing time is merely 0.05 ms on a GPU and 0.3 ms on a CPU platform, ensuring high computational efficiency that meets the requirements for online nuclear radiation monitoring and embedded real-time deployment. The proposed method offers a feasible technical solution for high-precision, highly robust real-time pulse shape discrimination in neutron-gamma mixed radiation fields.

## CRediT authorship contribution statement

**Shengduo Liu:** Conceptualization, Methodology, Software, Data curation, Validation, Investigation, Formal analysis, Writing-original draft. **Shiwei Jing:** Resources, Funding acquisition, Writing-review & editing, Conceptualization, Project administration. **Weiyang Zhang:** Software. **Jia Song**: Resources. **Sijia Zhou:** Formal analysis. **Hailong Xu:** Software, Visualization. **Yue Sun:** Software, Visualization. **Zebin Li:** Formal Analysis. **Yuxuan Gu:** Visualization. **Siqi Liu:** Formal analysis. **Tian Zhang:** Formal analysis. **Zhihua Gao:** Formal analysis. **Guofeng Qu:** Funding acquisition, Resources. **Fuquan Jia:** Formal analysis, Writing-review & editing.

## Acknowledgements

This research was supported by the Education Department of Jilin Province of

China (Grant No. JJKH20250303BS)